\documentstyle[pra,aps,twocolumn]{revtex}
\begin{document}
\flushbottom
\draft
\title{Creation of gap solitons in Bose-Einstein condensates}
\author{O.\ Zobay, E.\ M.\ Wright, and P.\ Meystre}
\address{Optical Sciences Center, University of Arizona, Tucson,
Arizona 85721}
\maketitle
\begin{abstract}
We discuss a method to launch gap soliton-like structures in
atomic Bose-Einstein condensates confined in optical traps. Bright vector
solitons consisting of a superposition of two hyperfine Zeeman sublevels
can be created for both attractive and repulsive interactions between
the atoms. Their formation relies on the dynamics of the atomic
internal ground states in two far-off resonant counterpropagating
$\sigma^+-\sigma^-$-polarized laser beams which form the optical trap.
Numerical simulations show that these solitons can be
prepared from a one-component state provided with an initial velocity.
\end{abstract}
\pacs{PACS numbers: 03.75.Fi, 05.30.Jp, 32.80.Pj}
\narrowtext
\section{Introduction}
The Gross-Pitaevskii equation (GPE) has been used
successfully in the recent past to explain various experiments on atomic
Bose-Einstein condensates (see, e.g., the references in
\cite{ScoBalBur97,ReiCla97}), and its validity
for the description of the condensate dynamics at zero
temperature is now well accepted. A further confirmation
would be provided by the observation of solitary matter waves, the
existence of which is generic to nonlinear Schr\"odinger wave equations such
as the GPE \cite{MorBalBur97}. Such solitary waves could also find
applications in the future, e.g., in the diffractionless transport of
condensates.

Various theoretical studies of this problem have already been performed,
predicting in particular the existence of bright solitons,
with corresponding spatially localized atomic density profiles,
for condensates with attractive interactions
\cite{ReiCla97}. Research on condensates with repulsive
interactions has focused on the formation of gray solitons which
correspond to dips in the atomic density. Their creation was
investigated in Refs.\ \cite{ScoBalBur97,DumCirLew97}, their general
properties were discussed in \cite{JacKavPet98}, and Ref.\
\cite{ReiCla97} worked out their analogy to the Josephson effect.

Complementary and previous to this work, the formation of atomic
solitons was also examined theoretically in the context of nonlinear
atom optics
\cite{LenMeyWri93,LenMeyWri94,SchLenMey94,ZhaWalSan94,DyrZhaSan97,HolAud97}.
In these studies the interaction between the atoms was assumed to result
from laser-induced dipole-dipole forces, but this theory has not been
experimentally tested so far.

The reliance on attractive interactions to achieve bright matter-wave solitons
in Bose-Einstein condensates is of course a serious limitation, due
to the difficulties associated with achieving condensation in the first
place for such interactions. The purpose of the present article is the
theoretical exposition of an experimentally realizable geometry
that allows one to create bright gap soliton-like
structures in Bose condensates, {\it for both attractive and repulsive
signs of the two-body scattering length.}  Gap solitons result from the
balance of nonlinearity and the effective
linear dispersion of a coupled system, e.g.,
counterpropagating waves in a grating structure, and appear
in the gaps associated with avoided crossings.
Gap solitons have previously been studied in a variety of physical
contexts, but particularly in nonlinear optics \cite{SteSip94}.
They were also studied in the framework of nonlinear atom optics
\cite{LenMeyWri94}, but in this case the two states involved are connected
by an optical transition, and the effects of spontaneous emission can
cause significant problems \cite{SchLenMey94}.

Several main reasons motivate our renewed interest in this problem.
First, we already mentioned that bright
gap solitons are known to exist in nonlinear systems irrespective of
whether the nonlinear interaction is repulsive or attractive
\cite{AceWab89}. With regard to atomic condensates this means that they
should be observable, at least in principle, also for Na and Rb where the
positive interatomic scattering length gives rise to a repulsive mean
interaction. Further, the study of bright solitary waves is of interest as
they might be easier to detect than gray ones, and they could find
future applications, e.g. in atomic interferometry \cite{HolAud97}.
An additional reason to
study atomic gap solitons is the fact that they consist inherently of a
superposition of two internal states, in our case two different Zeeman
sublevels of the atomic ground state. As such, they offer a
further example of a multicomponent Bose condensate the study of which
has already received much interest recently
\cite{HoShe96,GolMey97,MyaBurGhr97}. Finally, the recent demonstration
of far-off resonant dipole traps for condensates opens up the way to
the ``easy'' generation and manipulation of such spinor systems.

This paper is organized as follows. Section II describes our model. The
physics relevant for the generation of gap solitons as well as orders of
magnitudes for the various experimental parameters involved are discussed
in Sec.\ III, while Sec.\ IV presents a summary of our numerical results.
Finally, conclusions are given in Sec.\ V.
\section{The model}
The situation we consider for the generation of atomic gap solitons makes use
of the recently achieved confinement of Bose condensates in far off-resonant
optical dipole traps \cite{StaAndChi98}. We consider explicitly a trap
consisting of two focused laser beams of frequency $\omega_l$
counterpropagating in the $Z$-direction and with polarizations
$\sigma^+$ and $\sigma^-$, respectively. These lasers are used
to confine a Bose condensate which is assumed for concreteness to consist
of Na atoms. The condensate is initially prepared in the
$|g,F_g=1,M_g=-1\rangle$ atomic ground state. For lasers far detuned from
the resonance frequency $\omega_a$ of the nearest transition to an excited
hyperfine multiplet $|e,F_e,M_e=-F_e,\dots,F_e\rangle$ the dynamics of a
single atom in the trap can be described by an effective Hamiltonian of
the form \cite{Coh92}
\begin{eqnarray}\label{heff}
H_{eff}&=&\frac{{\bf P}^2}{2m}
+ d_0 \hbar\delta'({\bf R})|0\rangle \langle 0| \nonumber \\
&&+ d_1 \hbar\delta'({\bf R}) \left (|-1\rangle\langle -1|+
|1\rangle\langle 1|\right)
\\
&&+ d_2\hbar \delta' ({\bf R}) \left (|1\rangle\langle -1| e^{2iK_lZ}
+ |-1\rangle\langle 1| e^{-2iK_lZ}\right ), \nonumber 
\end{eqnarray}
which is derived by adiabatically eliminating the excited states in the
dipole and rotating wave approximations.
In the Hamiltonian (\ref{heff}), the operators ${\bf R}$
and ${\bf P}$ denote the center-of-mass position and momentum of the atom
of mass $m$, the ket $|j \rangle$ labels the magnetic sublevel of the
Na ground states, $|j \rangle \leftrightarrow |g,F_g=1,M_g=j\rangle$, and
$K_l=\omega_l/c$. Furthermore,
\begin{equation}
\delta'({\bf R})=\delta s({\bf R})/2  ,
\end{equation}
where we have introduced the detuning $\delta=\omega_l-\omega_a$ and the
position-dependent saturation parameter
\begin{equation}
s({\bf R})=\frac{{\cal D}^2 {\cal E}^2({\bf R})}{\delta^2+\Gamma^2/4}
\simeq \frac{{\cal D}^2 {\cal E}^2({\bf R})}{\delta^2}  .
\end{equation}
In this expression, ${\cal D}$ denotes the reduced dipole moment
between the states $|g\rangle$ and $|e\rangle$, $\Gamma$ is the
upper to lower state spontaneous emission rate, and ${\cal E}({\bf R})$
is the slowly varying laser field amplitude at point ${\bf R}$,
the plane-wave factors $\exp[i(\pm K_lZ-\omega_lt)]$ having already been
removed from the counterpropagating waves. In the following, we
assume that ${\cal E}$, which is identical for both fields,
varies only in the transverse $X$- and $Y$-directions
and is constant along the trap axis $Z$: This approximation is valid if
the longitudinal extension of the confined BEC is much less than the
Rayleigh range of the trapping fields,
a condition we assume is satisfied.  The
numerical coefficients $d_j$, which depend on the specific value of $F_e$,
are of the order of or somewhat less than unity. Note that
except insofar as $\Gamma$ appears in the saturation parameter
$s({\bf R})$,
the effects of spontaneous emission are neglected in this
description\footnote{In the discussion of the atomic dynamics and
Eq.\ (\ref{heff}) we have assumed that the initial state is coupled only
to one excited
hyperfine multiplet. However, in the optical trap the detuning of the
laser frequency is large compared even to the fine structure
splitting of the excited states, so that in principle several
different hyperfine multiplets should be taken into account.
Fortunately, the coupling to any of these multiplets gives rise to
addititional contributions to Eq.\ (\ref{heff}) which are of the same
analytical structure as the one given above. Only the values of $d_j$,
${\cal D}$, and $\Gamma$ are different. This means
that Eq.\ (\ref{heff}) may still be used in this case, the effects of
the additional multiplets being included as modifications of the values
of the coefficients $d_j$. For simplicity, however, we will use the
values of the $F_g=1$ $\to$ $F_e=1$ transition in the following,
i.e., $d_0=1/2$, $d_1=1/4$, and $d_2=-1/4$ \cite{Coh92}.}.

The first term in the single-particle Hamiltonian (\ref{heff})
describes the quantized center-of-mass atomic motion, the second and third
terms the (position dependent) light-shifts of the $j=0,\pm 1$ states,
and the final term, proportional to $d_2$, describes the coupling
between the $j=\pm 1$ states by the counterpropagating fields.
For example, coupling between the $j=-1$ and $j=1$ states arises from
the process involving absorption of one $\sigma^+$ photon and subsequent
re-emission of a $\sigma^-$ photon.  However, since the circularly
polarized fields are counterpropagating this process also involves
a transfer of linear momentum $2\hbar K_l$ along the $Z$-axis, and this
accounts for the appearance of the spatially periodic factors,
or gratings, $\exp(\pm2iK_lZ)$ in the coupling terms.  As shown below,
these gratings provide the effective linear dispersion which allows
for gap solitons in combination with the nonlinearity due to many-body
effects.

To describe the dynamics of the Bose condensate we introduce the
macroscopic wave function ${\bbox \Psi}({\bf R},t)=[\Psi_1({\bf R},t),
\Psi_{-1}({\bf R},t)]^T$ normalized to the total number of particles
$N$. Here $\Psi_0$ is omitted as it is coupled to
$\Psi_{\pm 1}$ neither by $H_{eff}$ nor by the nonlinearity if it
vanishes initially, which we assume in the following. The time
evolution of the spinor ${\bbox \Psi}({\bf R},t)$ is
determined by the two-component Gross-Pitaevskii equation
\begin{eqnarray}\label{gpe1}
i\hbar\frac{\partial {\bbox \Psi}}{\partial t}&=& H_{eff}
{\bbox \Psi}({\bf R},t) \\&+&
\left(\begin{array}{c} {[}U_a|\psi_1({\bf R},t)|^2 + U_b|\psi_{-1}
({\bf R},t)|^2{]}\psi_1({\bf R},t) \\  {[}U_b|\psi_1({\bf R},t)|^2 +
U_a|\psi_{-1}({\bf R},t)|^2{]}\psi_{-1}({\bf R},t)\nonumber
\end{array}\right)  .
\end{eqnarray}
In the following we approximate the nonlinearity coefficients by $U_a\approx
U_b=U=4\pi\hbar^2 a_{sc}/m$ with $a_{sc}$ the $s$-wave scattering length.

To identify the key physical parameters for gap soliton formation, and to
facilitate numerical simulations, it is convenient to
re-express Eq.\ (\ref{gpe1}) in a dimensionless form by introducing scaled
variables $\tau=t/t_{c}$, ${\bf r}={\bf R}/l_{c}$ and
$\psi_j=\Psi_j/\sqrt{\rho_{c}}$ with
\begin{eqnarray}
t_{c}&=&1/(d_2\delta'_0), \label{tchar}\\
l_{c}&=&t_{c}\cdot \hbar K/m,\label{lchar}\\
\rho_{c}&=&|d_2\hbar\delta'_0/U|,\label{nchar}  ,
\end{eqnarray}
where $\delta'_0=\delta'({\bf R}=0)$.
Note that for our choice of $d_2=-1/4$, and
for red detuning, we have $d_2\delta'_0>0$. Equation (\ref{gpe1}) then reads
\begin{eqnarray}\label{gpe2}
i\frac{\partial {\bbox \psi}}{\partial \tau}&=&\left[\begin{array}{cc}
-M\Delta + \frac{d_1\delta'({\bf r})}{d_2\delta'_0} & e^{2ik_lz}
\delta'({\bf r})/\delta'_0 \\ e^{-2ik_lz}\delta'({\bf r})/\delta'_0 &
-M\Delta + \frac{d_1\delta'({\bf r})}{d_2\delta'_0} \end{array}
\right]\left(\begin{array}{c} \psi_1 \\ \psi_{-1} \end{array} \right)
\nonumber \\
& & +\left[\mbox{sgn}(d_2\hbar\delta'_0/U) (|\psi_1|^2
+|\psi_{-1}|^2) \right]\left(\begin{array}{c} \psi_1 \\
\psi_{-1} \end{array} \right)  ,
\end{eqnarray}
where $\Delta$ is the Laplacian in scaled variables, and we have introduced
the dimensionless mass-related parameter
\begin{equation}
M=d_2\delta'_0 m/(2\hbar K_l^2)  ,
\end{equation}
so that $k_l=K_l l_{c}=1/(2M)$.
\section{Gap solitons}
In this section we discuss the conditions under which Eqs.\ (\ref{gpe2})
yield gap soliton solutions.  Rather than reproducing the explicit
analytic forms of these solutions, which are readily available in
the literature, here we introduce the reduced equations which
yield gap solitons, and discuss the physics underlying their
formation.  Estimates for the orders of magnitude of various parameters
characterizing atomic gap solitons are also given.
\subsection{Reduced soliton equations}
Two key approximations underly the appearance of gap solitons: First,
we neglect all transverse variations of the electromagnetic and
atomic fields, thereby reducing the problem to one spatial variable
$z$.  Furthermore, we can set $\delta'({\bf r})/\delta'_0=1$.
Second we express the atomic fields in the form
\begin{equation}\label{trans}
\psi_{\pm 1}(z,t)=\exp\{i[\pm k_l z-(1/(4M)-1)\tau]\}\phi_{\pm 1}(z,t)  ,
\end{equation}
and we assume that
the atomic field envelopes $\psi_{\pm 1}(z,t)$ vary slowly in
space in comparison to the plane-wave factors that have been
separated out, so that only first-order spatial derivatives
of the field envelopes need be retained and only the spatial
harmonics indicated included.  Under these assumptions
Eqs.\ (\ref{gpe2}) reduce to
\begin{eqnarray}\label{gpe3}
&&i\left (\frac{\partial}{\partial\tau}\pm 2k_l
\frac{\partial}{\partial z} \right )
\left(\begin{array}{c} \phi_1 \\
\phi_{-1}\end{array} \right)=\left(\begin{array}{cc} 0 & 1\\
1 & 0 \end{array}\right)
\left(\begin{array}{c} \phi_1 \\ \phi_{-1} \end{array} \right)
\nonumber \\
&&+\mbox{sgn}(d_2\hbar\delta'_0/U) (|\phi_1|^2
+|\phi_{-1}|^2)\left(\begin{array}{c} \phi_1 \\
\phi_{-1} \end{array} \right)  .
\end{eqnarray}
Aceves and Wabnitz \cite{AceWab89} have shown that these dimensionless
equations have explicit
travelling solitary wave solutions of hyperbolic secant form.
Thus, the optical trapping geometry we propose here can
support atomic gap solitons under the appropriate conditions.

Having established that our system can support gap solitons
our goal in the remainder of this paper is to demonstrate through
numerical simulations that these solitons, or at least a remnant
of them, can arise for
realistic atomic properties and that they can be created from
physically reasonable initial conditions.  In particular,
the exact gap solitons solutions are coherent superpositions
of the $j=\pm 1$ states where the phase and amplitude of the
superposition varies spatially in a specific manner: it is
not a priori clear that these gap solitons can be accessed
from an initial state purely in the $j=-1$ state for example.
Furthermore, inclusion of transverse variations and spatial
derivatives beyond the slowly-varying envelope approximation
introduced above could, in principle, destroy the solitons
\cite{ChaMalFri98}.
For the numerical simulations to be presented here we work
directly with Eqs. (\ref{gpe2}) which does not invoke these
approximations.
\subsection{Intuitive soliton picture}
A simple and intuitively appealing explanation of the reason
why Eq.\ (\ref{gpe2}) supports soliton solutions goes as follows:
Consider first the one-di\-men\-sio\-nal nonlinear Schr\"odinger equation
\begin{equation}
i\dot\psi=-M\partial^2\psi/\partial z^2+g|\psi|^2\psi  .
\end{equation}
This equation has {\it bright} soliton solutions if the effects of dispersion
and nonlinearity can cancel each other. For this to happen, it is
necessary that $Mg<0$. In the usual case the mass-related coefficient
$M$ is positive, so that bright solitons can only exist in condensates with
attractive interactions $g<0$. But consider now the dispersion relation
$\omega(k)=Mk^2\pm\sqrt{1+k^2}$ for the linear part of Eq.\ (\ref{gpe2})
obtained after neglecting the transverse dimensions and performing
the transformation
\begin{equation}\label{transform}
\psi_{\pm 1}=a_{\pm 1}\exp\{i[k_{\pm 1}z-\omega(k)\tau]\}.
\end{equation}
Thereby, $k_{\pm 1}=k\pm k_l$ for the $j=\pm 1$ states, and $k$ is a
relative longitudinal wave vector. The dispersion relation
consists of two branches, which in the absence of linear
coupling take the form of two parabolas corresponding to the free
dynamics of the internal states $|\pm 1\rangle$. However, the linear
coupling between these states results in an avoided crossing
at $k=0$, see Fig.\ \ref{fig1}. If the system is in a superposition of
eigenstates pertaining to the lower branch of the dispersion relation, then
at the crossing it can be ascribed a {\it negative} effective mass. One can
thus expect that in this case the system can support soliton solutions
even though the interaction is repulsive. For an attractive interaction,
in contrast, soliton creation should be possible in all regions of the
spectrum with positive effective mass.

From the dispersion relation picture, one can easily infer further
properties of repulsive interaction solitons. First, they will only exist
for weak enough dispersion, as the lower branch of the dispersion curve
has a region with negative curvature only as long as the dimensionless
mass $M<0.5$. Also,
the maximum possible velocity can be estimated to be of the order of
$[1-(2M)^{2/3}]^{3/2}$ which is the group velocity at
the points $\pm k_0=\pm \sqrt{(2M)^{-2/3}-1}$
of vanishing curvature in the dispersion relation. Finally, for a soliton
at rest the contributions of the internal states
$|1\rangle$ and $|-1\rangle$
will approximately be equal, but solitons traveling with increasing
positive, resp. negative, velocity will be increasingly dominated by the
$|1\rangle$, resp.\ $|-1\rangle$, contribution.

This qualitative discussion is in agreement with the analytic
results of Ref.\cite{AceWab89}.
More precisely, the solutions of Ref. \cite{AceWab89} are
solitary waves. In the following, we will be concerned with the creation
of long-lived localized wave packet structures which are brought about
by the interplay between nonlinearity and dispersion described above. We
will continue to refer to these structures as gap solitons for simplicity.
\subsection{Soliton estimates}
We now turn to a discussion of the typical orders of magnitude which
characterize the soliton solutions of Eq.\ (\ref{gpe1}). We note
from the outset that the analytical solutions of Ref.\ \cite{AceWab89},
as well as our numerical simulations, indicate that these characteristic
scales can be directly inferred from the scale variables in
Eqs.\ (\ref{tchar})--(\ref{nchar}) which bring the Gross-Pitaevskii
equations into dimensionless form.
For example, the spatial extension of the scaled wave function
${\bbox \psi}$, as well as the total norm
$\int dv(|\psi_1|^2 + |\psi_{-1}|^2)$ are of order 
unity.\footnote{Note that the precise values of $d_1$ and $d_2$ are only of
relevance for the scaling between Eqs.\ (\ref{gpe1}) and (\ref{gpe2}).
They do not influence the essential physics of the system.}

In order to obtain estimates for the characteristic length, time and
density introduced in Eqs.\ (\ref{tchar})--(\ref{nchar}) we use the
parameter values of the  Na experiment of Ref. \cite{StaAndChi98} as a
guidance. For Sodium, $\Gamma=61$ MHz, the saturation intensity
$I_s=6.2$ mW/cm$^2$ and the resonance wavelength $\lambda_a=589$ nm.
Choosing the trap wavelength to be far red-detuned from this value, with
$\lambda_l=985$ nm, and a maximum laser intensity
$I\approx 1$ kW/cm$^2$ one obtains a characteristic
scale $t_{c}$ for the time evolution of the condensate of
the order of 50 $\mu$s. The characteristic length is obtained by
multiplication with the recoil velocity $v_{rec}$ =1.8 cm/s, which
yields $l_{c}\approx 1 - 2$ $\mu$m. This yields the dimensionless mass
$M \approx 0.1$. Finally, the order of
magnitude of the characteristic density $\rho_{c} =10^{14}$
cm$^{-3}$, which means that a soliton typically contains of the order of
$\rho_{c} l_{c}^3\approx 100 - 1000$ atoms.
These estimates are confirmed by our numerical simulations, which
show that the typical extension of a soliton is several $l_{c}$ in the
$z$-direction, about one $l_{c}$ in the transverse direction and it
contains about 1000 atoms.

Our numerical simulations show that the maximum dimensionless atomic
density in a soliton is always of the order of $\rho_m=0.1$, which
appears to produce the nonlinearity necessary to balance the effects of
dispersion. From this value, it is possible to obtain a first estimate
of the transverse
confinement of the condensate required in our two-dimensional model:
We assume that the transverse spatial dependence of the
atomic density can be modeled as the normalized ground state density
$g(X)$ of the harmonic trap potential $\delta'(X)= m\omega_x^2 X^2/2$
\cite{JacKavPet98}. The soliton density can hence be roughly estimated as
$\rho({\bf R})=\Theta(Z)\Theta(\Delta Z -Z)g(X)N/\Delta Z$, where
$\Theta(Z)$is the Heaviside step function, $N$ a typical total number of
atoms in the soliton and $\Delta Z$ its length. From this condition, and 
using the typical values for $N$ and $\Delta Z$ previoulsy discussed leads 
to a lower limit for $\omega_x$ in the range between 100 and 1000 Hz. 
Altogether, these various estimates are well within experimental reach.

\section{Numerical results}

Having characterized the idealized
gap soliton solutions of Eq.\ (\ref{gpe1}) we
now investigate whether they can be accessed from realistic
initial conditions. To this end, we study
numerically the following situation. A condensate of $N_0$ atoms in the 
internal state $|-1\rangle$ is initially prepared in a conventional 
optical dipole trap which provides only a tranverse
confinement potential $V_c$. This potential is assumed to be Gaussian,
with a trapping frequency $\omega_x$ at the bottom. Axially, the
condensate is confined by a harmonic magnetic trap of frequency
$\omega_z$. At $t=0$ the magnetic trap is turned off and the polarizations
of the trapping light fields are switched to the $\sigma^+ - \sigma^-$ 
configuration, with $V_c$ being unchanged.

The simulations are performed in two spatial dimensions only, $x$ and
$z$, as this can already be expected to capture the relevant physics
without requiring excessive computational resources.
We concentrate on the more
interesting case of a condensate with repulsive interactions since
that is the new case in which solitons are expected.
In order to estimate atom numbers the transformation between Eqs.\
(\ref{gpe2}) and (\ref{gpe1}) is performed after replacing $U$ by
$U/\sqrt{\langle X^2 \rangle}$ where the variance $\langle X^2 \rangle$
is determined from the two-dimensional wave packet structure in
question.

The main purpose of the numerical simulations is to show that gap
solitons can be formed out of condensate wave functions whose
initial parameters lie within a relatively broad range. It is only
necessary to choose $\omega_x$, $\omega_z$, and $N_0$ such that the
spatial extension of the initial condensate and the atom number are
comparable to typical soliton values. They need not take on precisely
defined values and the initial wave function does not have to match
closely the form of a soliton.

However, the condensate will not couple effectively to a soliton if it
is at rest initially. Such a situation corresponds to the point
with $k=k_l$ in Fig.\ \ref{fig1} where the effective mass is still
positive ($k_l>k_0$). The key to an efficient generation of solitons is
therefore to provide the condensate with an initial velocity $V_{in}=\hbar
K_{in}/m$ close to the recoil velocity $V_r$ in the $Z$-direction.
This may be achieved, e.g., by suddenly displacing the center of the
magnetic trap. The initial wave function can
then be written as $\psi_{-1}(x,z,0)=\psi_g(x,z,0) \exp(i k_{in}z)$
with $\psi_g(x,z,0)$ the ground state of the combined optical and magnetic
trap and $k_{in}=K_{in} l_{c}$ \cite{MorBalBur97}.
It is thus placed in the vicinity of the avoided crossing.
Experimentally, condensates have already been accelerated to velocities in
this range by a similar method in connection with the excitation of
dipole oscillations \cite{DurKet98}.

Figure \ref{fig2} shows an illustrative example for the formation of a
soliton out of the initial distribution. It depicts the evolution of the
transverse averaged atomic density
\begin{equation}
N(z;\tau)=\int dx (|\psi_1(x,z;\tau)|^2 +|\psi_{-1}
(x,z;\tau)|^2)  ,
\end{equation}
as a function of the scaled variables $z$ and $\tau$. In
this example, $\lambda_l=985$ nm, the maximum intensity $I=0.88$ kW/cm$^2$,
$\omega_x=6000$ s$^{-1}$, $\omega_z=5100$ s$^{-1}$, and $V_{in}=0.9V_r$. 
The characteristic scales are thus $l_{c}=1.4$ $\mu$m and $t_{c}=77$
$\mu$s, the coefficient $M=0.06$. We choose
$\int dxdz|\psi_{-1}(x,z,0)|^2=6.21$
which corresponds to an
initial atom number of 2900, approximately. Figure \ref{fig2} shows the
formation of a soliton after an initial transient phase having a
duration of 50 $t_{char}$, approximately. This transient phase is
characterized by strong ``radiation losses''. They occur because half of
the initial state pertains to the upper branch of the
dispersion relation $\omega(k)$ which cannot sustain solitons. The shape
of the created soliton is not stationary in time but appears to
oscillate. Further examination shows that norm of the $j=-1$ state
is slightly larger than that of the $j=1$ state, as is expected for a soliton
moving slowly in negative $z$-direction \cite{AceWab89}.
Figure \ref{fig3} shows the
atomic density $P(x,z)=|\psi_1(x,z;\tau)|^2+|\psi_{-1}(x,z;\tau)|^2$ in
the soliton at $\tau=100$. The soliton contains about 500 atoms. The inset
depicts the longitudinally integrated density $\int dz (|\psi_1(x,z;\tau)|^2
+|\psi_{-1}(x,z;\tau)|^2)$ and the transverse confinement potential in
the shape of an inverted Gaussian. The soliton thus spreads out over
half the width of the potential well, approximately.

Various numerical simulations were performed in order to assess the
dependence of soliton formation on the various initial parameters.
When changing the atom number $N_0$ a soliton was formed over
the whole investigated range between 500 and 4000 atoms. With increasing
$N_0$ and thus increasing effects of the nonlinearity the final velocity
of the soliton changed from negative to positive values. For large $N_0$
a tendency to form two soliton wave packets out of the initial state was
observed, however, the formation of each of these solitons is accompanied
by large radiation losses which destabilize the other one. As to the
transverse confinement parameter $\omega_x$ soliton formation was
observed for $\omega_x \geq 4000$ s$^{-1}$. At $\omega_x=2000$ s$^{-1}$
a stable structure was no longer attained which is in rough agreement
with the estimate given above. Whereas these numbers indicate a
relatively large freedom in the choice of $N_0$ and $\omega_x$ (for
$\omega_z$ similar results can be expected) a somewhat more restrictive
condition is placed on the initial velocity $V_{in}$. Its value should
be chosen from the interval between 0.8 and 1.0 $V_r$ in order to
guarantee soliton formation, the lower bound being determined by
the point of vanishing curvature in the dispersion relation. For
$V_{in}> V_r$ the initial wave function is situated more and more on the
upper branch of the dispersion relation so that the tendency to form
solitons is diminished rapidly.

\section{Summary and conclusion}

In conclusion, we have demonstrated that gap soliton-like structures can be
created in a Bose condensate confined in an optical dipole trap formed
by two counterpopagating $\sigma^+-\sigma^-$-polarized laser beams.
Bright solitons can be formed not only for atomic species with
attractive interactions but also in the repulsive case. This is
rendered possible because the atoms can be ascribed a negative effective
mass if their velocity is close to the recoil velocity. The repulsive
interaction solitons are inherently superpositions of two hyperfine
Zeeman sublevels. The discussion of characteristic scales and numerical
simulations indicated that the actual observation of these structures 
should be achievable within the realm of current experimental possibilities.

In our theoretical treatment spontaneous emission was
neglected, an approximation justified by the large detunings in the optical
trap \cite{StaAndChi98}. The effects of anti-resonant terms, which were
also ignored, might be of more importance. This question,
as well as three-dimensional numerical studies, could be the subject of
future work.

\acknowledgements
We have benefited from numerous discussions with E.\ V.\ Goldstein.
This work is supported in part by the U.S. Office of Naval Research
Contract No.\ 14-91-J1205, by the National Science Foundation Grant
PHY95-07639, by the U.S. Army Research Office and by the
Joint Services Optics Program.

\begin{figure}
\caption{Dispersion relation of the linear part of Eq.\ (\protect\ref{gpe2})
for the parameter value $M=0.06$.The $k$-value associated with a
condensate at rest is denoted by $k_l$, the point of vanishing curvature
is indicated by $k_0$.}
\label{fig1}
\end{figure}

\begin{figure}
\caption{Formation of a soliton out of an initial distribution. Depicted
is the integrated density $N(z;\tau)=\int dx $$(|\psi_1(x,z;\tau)|^2
+|\psi_{-1}(x,z;\tau)|^2)$ in scaled variables. Parameter values given
in the text.}
\label{fig2}
\end{figure}

\begin{figure}
\caption{Atomic density $P(x,z)=$ $|\psi_1(x,z;\tau)|^2+$ 
$|\psi_{-1}(x,z;\tau)|^2$
of the soliton of Fig.\ \protect\ref{fig2} at $\tau=100$. The inset
shows the integrated density $Q(x)=\int dz (|\psi_1(x,z;\tau)|^2
+|\psi_{-1}(x,z;\tau)|^2)$ and the transverse confinement potential
$V(x)$.}
\label{fig3}
\end{figure}

\end{document}